\documentclass[11pt]{article}
\usepackage{moriond,epsfig}

\def\be{\begin{equation}}
\def\ee{\end{equation}}
\def\bea{\begin{eqnarray}}
\def\eea{\end{eqnarray}}

\begin{document}
\vspace*{4cm}
\title{MINIMAL REALISTIC $SU(5)$ SCENARIO}

\author{Ilja Dor\v{s}ner}

\address{The Abdus Salam
International Centre for Theoretical Physics\\
Strada Costiera 11, 34014 Trieste, Italy}

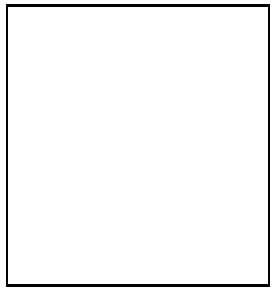
\maketitle

\abstracts{We present phenomenological aspects of the simplest
realistic $SU(5)$ grand unified theory---the theory with the ${\bf
5}$, ${\bf 15}$, and ${\bf 24}$ dimensional representations in the
Higgs sector. We show that a successful gauge coupling unification
sets experimentally accessible upper bound on the total proton
decay lifetime. It also relates proton decay lifetime to scalar
leptoquark mass in an experimentally testable manner. We also
discuss an addition of gauge singlets---both fermions and
bosons---to the simplest scenario and comment on relevant
phenomenological consequences of such modifications.}

\section{Introduction}

Grand unified theories (GUTs) are considered to be the most viable
candidates for the physics beyond the Standard Model. Through
matter unification and unification of strong and electroweak
interactions they always generate two predictions: (1) gauge
couplings unify and (2) proton decays. Testing the first
prediction is practically impossible since unification takes place
at energy scales beyond our reach. However, the second one can be
experimentally investigated thereby offering a very promising way
to test grand unification. It is thus important to single out and
investigate grand unified theories where proton decay is both
accurately predicted and experimentally reachable.

We have recently proposed~\cite{Dorsner:2005fq,Dorsner:2005ii} the
simplest realistic GUT model with both of those properties. It
comes in a form of a particularly simple extension of the
well-known Georgi-Glashow $SU(5)$ model~\cite{GG}. In particular,
the scenario contains not only the usual three generations of
matter fields and the ${\bf 5}$ and ${\bf 24}$ dimensional Higgs
representations, but also one ${\bf 15}$ dimensional Higgs. And,
it includes \textit{all}\/ possible $SU(5)$ invariant operators.

Despite a large number of parameters in the Lagrangian the
scenario is still very predictive. This predictivity is primarily
attributed to the simplicity of the Higgs sector. But, two generic
features of non-supersymmetric $SU(5)$ framework also boost
predictivity. Firstly, $SU(5)$ is the only simple group with the
Standard Model (SM) embedding that has unique single step symmetry
breaking. This allows for an accurate determination of the unified
scale---the so-called GUT scale $M_{GUT}$. Secondly, the least
model dependent and usually dominant contribution to proton decay
comes from an exchange of only one set of superheavy gauge bosons
with mass $M_V$. Clearly, if $M_V$ is taken to define to the GUT
scale the former property also implies accurate prediction for
proton decay. That is exactly what happens in our case.

In the following, we briefly present the simplest realistic
$SU(5)$ theory. Again, even though the proposed scenario has
uncorrelated regions in the Yukawa sector, the simplicity of its
Higgs sector guarantees both its testability and refutability in
near future.

\section{The simplest $SU(5)$ scenario}

The scenario we propose has a following particle content of the
Higgs sector\cite{Dorsner:2005fq}: ${\bf 5}= \Psi=(\Psi_D,
\Psi_T)=({\bf 1},{\bf 2},1/2)+({\bf 3},{\bf 1},-1/3)$, ${\bf
15}=\Phi =(\Phi_a, \Phi_b, \Phi_c)=({\bf 1},{\bf 3},1)+({\bf
3},{\bf 2},1/6)+({\bf 6},{\bf 1},-2/3)$ and ${\bf 24}= \Sigma
=(\Sigma_8, \Sigma_3, \Sigma_{(3,2)}, \Sigma_{(\bar{3}, 2)},
\Sigma_{24})=({\bf 8},{\bf 1},0)+({\bf 1},{\bf 3},0)+({\bf 3},{\bf
2},-5/6) +(\overline{{\bf 3}},{\bf 2},5/6)+({\bf 1},{\bf 1},0)$,
where we use the SM ($SU(3) \times SU(2) \times U(1)$)
decomposition to set our notation. $\Sigma_{(3,2)}$ and
$\Sigma_{(\bar{3}, 2)}$ are fields eaten by the superheavy gauge
fields $V$. We define the GUT scale through their common mass,
i.e., we set $M_V=M_{GUT}$. As always, a vacuum expectation value
(VEV) of $\Sigma_{24}$ breaks $SU(5)$ while VEV of $\Psi_D$
triggers the SM symmetry breaking. During the latter stage
$\Phi_a$ develops an induced VEV that eventually yields neutrino
mass through the type II see-saw mechanism~\cite{TypeII}.

On renormalizable level our scenario predicts $Y_D = Y_E^T$ at the
GUT scale, where $Y_{D(E)}$ is the down quark (charged lepton)
Yukawa coupling matrix. This prediction however disagrees with
experiments, especially in the case of the first and second
generation. In order to correct that we include higher-dimensional
$SU(5)$ invariant operators~\cite{Ellis:1979fg}. (If we demand
renormalizability we must introduce $\bf{45}$ dimensional Higgs
representation~\cite{Georgi:1979df}. However, falsifiability of
such an extension is not guaranteed~\cite{Dorsner:2006dj}.)

Clearly, our scenario accommodates realistic fermionic mass
spectrum. What remains to be investigated is the issue of gauge
unification and its compatibility with proton decay constraints.

\subsection{Unification vs.\ proton decay}

There are four masses---$M_{GUT}$, $M_{\Sigma_3}$, $M_{\Phi_a}$
and $M_{\Phi_b}$---and two equations that govern gauge coupling
unification at the one-loop level~\cite{Dorsner:2005fq} in our
case. Actually, there are three renormalization group
equations---one for each gauge coupling of the SM. However,
elimination of the unified coupling constant $\alpha_{GUT}$ leaves
only two relevant equations~\cite{Giveon:1991zm}. These are
\begin{equation}
\label{condition1} \frac{B_{23}}{B_{12}}=0.719\pm0.005,\qquad
\rm{and} \qquad\ln \frac{M_{GUT}}{M_Z}=\frac{184.9 \pm
0.2}{B_{12}},
\end{equation}
where the right-hand sides reflect the latest experimental
measurements of the SM parameters~\cite{Eidelman:2004wy}. The
left-hand sides depend on particular mass spectrum of the particle
content of the theory at hand. More precisely, $B_{ij}=B_i - B_j$,
where $B_i$ coefficients are given by:
\begin{equation}
\label{r} B_i = b_i+\sum_{I} b_{iI} r_{I}, \qquad r_I=\frac{\ln
M_{GUT}/M_{I}}{\ln M_{GUT}/M_{Z}}.
\end{equation}
$b_i$ are the SM coefficients while $b_{iI}$ are the one-loop
coefficients of any additional particle $I$ of mass $M_I$ ($M_Z
\leq M_I \leq M_{GUT}$). (Recall, for the case of $n$ light Higgs
doublet fields $b_1=40/10+n/10$, $b_2=-20/6+n/6$ and $b_3=-7$.)
The $B_{ij}$-coefficient contributions in our scenario are listed
in Table~\ref{tab:table1}. Note that the SM case yields
$B_{23}/B_{12}=0.53$. This means that additional particles with
intermediate masses $M_I$ are required for successful unification.
In our case these particles are clearly $\Sigma_3$, $\Phi_a$ and
$\Phi_b$.
\begin{table}[h]
\caption{\label{tab:table1} $B_{ij}$ coefficients.}
\begin{center}
\begin{tabular}{|lccccccccc|}
\hline
     &Higgsless SM&$\Psi_D$&$\Psi_T$ & $V$ & $\Sigma_8$
     & $\Sigma_3$ & $\Phi_a$ & $\Phi_b$ & $\Phi_c$\\
\hline $B_{23}$& $\frac{11}{3}$&$\frac{1}{6}$&$-\frac{1}{6}
r_{\Psi_T}$ &$-\frac{7}{2}r_V$ &$-\frac{1}{2}
r_{\Sigma_8}$&$\frac{1}{3} r_{\Sigma_3}$ &$\frac{2}{3}r_{\Phi_a}$
&$\frac{1}{6} r_{\Phi_b}$ &$-\frac{5}{6} r_{\Phi_c}$\\
$B_{12}$&$\frac{22}{3}$&$-\frac{1}{15}$&$\frac{1}{15} r_{\Psi_T}$
&$-7r_V$ &0 &$-\frac{1}{3} r_{\Sigma_3}$
&$-\frac{1}{15}r_{\Phi_a}$ &$-\frac{7}{15} r_{\Phi_b}$
&$\frac{8}{15} r_{\Phi_c}$\\
\hline
\end{tabular}
\end{center}
\end{table}

\begin{figure}[h]
\begin{center}
\includegraphics[width=4in]{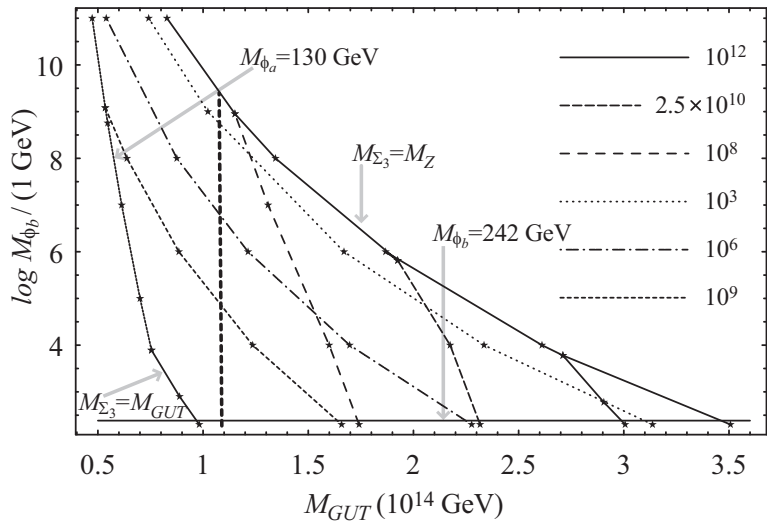}
\end{center}
\caption{\label{figure:2} Unification of the gauge couplings at
the two-loop level. Stars correspond to exact numerical two-loop
unification solutions. There are two sets of lines of constant
value. The steeper set is associated with $M_{\Phi_a}$ and the
other one represents the lines of constant $M_{\Sigma_3}$. All
masses are in GeV units. The region to the left of the vertical
dashed line is excluded by the proton decay experiments.}
\end{figure}

In this paper we present the outcome of the two-loop level
unification analysis in terms of $M_{\Sigma_3}$ and $M_{\Phi_a}$
contours in the $M_{GUT}$--$M_{\Phi_b}$ plane in
Fig.~\ref{figure:2}. Stars represent points that correspond to
exact numerical unification while lines represent linear
interpolation.

The sail-like region in Fig.~\ref{figure:2} represents the viable
parameter space under the assumption that $\Psi_T$, $\Sigma_8$ and
$\Phi_c$ are at or above the GUT scale. It is bounded from the
left and below by experimental limits on $M_{\Phi_a}$ and
$M_{\Phi_b}$, respectively. The right bound stems from the
requirement that $M_{\Sigma_3} \geq M_Z$. Note that $\Phi_b$ is a
scalar leptoquark and hence very interesting generator of new
physics~\cite{Dorsner:2005fq,Dorsner:2005ii}. For example, Large
Hadron Collider (LHC) aims to place more stringent lower limits on
the mass of $\Phi_b$ at $1$\,TeV.

The region to the left of the vertical thick dashed line in
Fig.~\ref{figure:2} is excluded by the present limits on the
proton decay lifetime. In order to generate this bound we
appropriately take the case of maximal flavor suppression of the
gauge $d=6$ proton decay operators~\cite{Dorsner:2004xa} and use
$\alpha=0.015\,\textrm{GeV}^3$ for the value of nucleon matrix
element~\cite{Aoki:2004xe}. Experimental limit we take as input
reads $\tau_p(p \rightarrow \pi^0 e^+)>5.0 \times
10^{33}$\,years~\cite{Eidelman:2004wy}.

What happens if we relax the $M_{\Psi_T},M_{\Sigma_8},M_{\Phi_c}
\geq M_{GUT}$ assumption? As one lowers $M_{\Psi_T}$ and
$M_{\Phi_c}$ the $M_{\Phi_a}=130$\,GeV line in Fig.~\ref{figure:2}
moves slowly to the left while, at the same time, the
$M_{\Sigma_3}=M_Z$ line moves very rapidly in the same direction
until the allowed region shrinks to a point. Hence, any scenario
in which $M_{\Psi_T}$ or $M_{\Phi_c}$ or both are below $M_{GUT}$
would be more exposed to the tests through the proton decay
lifetime measurements and accelerator searches than the scenario
shown in Fig.~\ref{figure:2}.

If, on the other hand, one lowers the mass of $\Sigma_8$, the
$M_{\Phi_a}=130$\,GeV line moves to the right more rapidly than
the $M_{\Sigma_3}=M_Z$ line until the allowed region becomes a
point when $M_{\Sigma_8}$ reaches $M_Z$. At that point $M_{GUT}$
is $4.6 \times 10^{14}$\,GeV for $M_{\Sigma_3}=M_Z$,
$M_{\Phi_a}=6.4 \times 10^3$\,GeV, $M_{\Phi_b}=242$\,GeV and
$\alpha_{GUT}^{-1}=37.06$. We use these particular values to
derive an accurate upper bound on the proton decay lifetime:
\begin{equation}
\tau_p^{(\textrm{two-loop})}  \leq 1.4 \times
10^{36}\,\textrm{years}.
\end{equation}
This bound follows from a more general model independent
inequality~\cite{Dorsner:2004xa}:
\begin{equation}
\tau_p \leq 6 \times 10^{39} \alpha_{GUT}^{-2} (M_V/10^{16}
\textrm{GeV})^4 (0.003 \textrm{GeV}^3/\alpha)^2\,\rm{years},
\end{equation}
which is applicable to any simple group with the SM embedding.
(Note, in the case of scenarios with partial gauge coupling
unification such as the Pati-Salam~\cite{Pati:1974yy} or flipped
$SU(5)$~\cite{DeRujula:1980qc,Georgi:1980pw,Barr:1981qv} scenario
the proton can be stable~\cite{Dorsner:2004jj}.)

Clearly, an improvement in the proton lifetime measurements by a
$4^4$ factor is called for to completely rule out this GUT
scenario. The situation is actually even more promising; even a
mild improvement in the proton lifetime bounds (by a factor of
fifteen) would make our scenario incompatible with exact
unification unless either $\Phi_b$ or $\Sigma_3$ resides below
$10^3$\,GeV. (This certainly makes them accessible in accelerator
experiments.) The next generation of proton decay experiments aims
at improving lower bounds on partial lifetimes by a few orders of
magnitude. For instance, the goal of Hyper-Kamiokande is to
explore the proton lifetime at least up to $\tau_p/B(p \to e^+
\pi^0) > 10^{35}$ years and $\tau_p/B(p \to K^+ \bar{\nu})>
10^{34}$ years in about 10 years~\cite{Nakamura:2003hk}. Thus, our
minimal GUT scenario will be tested and/or ruled out at the next
generation of proton decay and accelerator experiments.

\section{Adding singlets}

Let us comment on possible additions of $SU(5)$ singlets---both
spin $1/2$ and spin $0$ particles---to our scenario.

As we argued, our scenario is tailor made to generate neutrino
masses through the Type II seesaw mechanism~\cite{TypeII}. But,
one can also introduce right-handed neutrinos---singlets of
$SU(5)$---and generate additional neutrino mass contribution
through the so-called Type I seesaw mechanism~\cite{seesaw}. That
sort of addition obviously cannot change our discussion on the
gauge coupling unification and related constraints---at least at
the one-loop level. Hence, there is no need to modify our previous
discussion on proton decay. However, the addition of right-handed
neutrinos has potential to accommodate viable mechanism that
explains the origin of the baryon asymmetry observed in the
universe~\cite{Spergel:2006hy}.

Among viable mechanisms to explain primordial matter-antimatter
asymmetry, leptogenesis~\cite{Fukugita:1986hr} has undoubtedly
become one of the most compelling scenarios. In our case,
leptogenesis could proceed through the decay of the lightest
triplet scalar $\Phi_a$. As has been shown the scalar triplet with
a mass $M_{\Phi_a} \geq 10^{9-10}$\,GeV~\cite{Hambye:2005tk}
constitutes a natural candidate for a successful triplet
leptogenesis if our minimal $SU(5)$ scenario is extended with at
least one singlet fermion field. Combining the above leptogenesis
bounds with the ones shown in Fig.~\ref{figure:2}, we observe that
successful triplet leptogenesis excludes the region of the
parameter space where the mass of the triplet $M_{\Phi_a}$ is
below $10^9$--$10^{10}$\,GeV. This in turn would imply that
leptoquark $\Phi_b$ could be light enough ($M_{\Phi_b} <
10^{6-7}$\,GeV) to open the possibility of direct production at
the next generation of collider experiments. It should be noted
however that in general if both Type I and Type II seesaws are
present leptogenesis could proceed in a way that avoids generation
of any bounds on the mass spectrum of relevant
particles~\cite{Hambye:2003ka,Hambye}.

Let us also comment on a possible addition of scalar $SU(5)$
singlet fields from point of view of a successful inflationary
scenario~\cite{Guth:1980zm,Linde:1981mu,Albrecht:1982wi}. Namely,
in view of the latest three-year WMAP data~\cite{Spergel:2006hy}
there has been a renewed interest in a class of inflationary
models based on a quartic Coleman-Weinberg potential where a gauge
singlet scalar field plays iflaton role~\cite{Shafi:1983bd}. It
has been recently shown~\cite{Shafi:2006cs} that such a class of
models can be in good agreement with WMAP
data~\cite{Spergel:2006hy}. This conclusion also applies to our
scenario if one additional $SU(5)$ singlet scalar field is
included.

Finally, another $SU(5)$ scalar singlet can also be introduced to
address the existence of Dark Matter~\cite{Spergel:2006hy}. This
approach however would require additional discreet
symmetries~\cite{McDonald:1993ex} in order to forbid certain
couplings.
\section{Additional comments}
\begin{itemize}
\item Higher-dimensional operators help us generate realistic mass
spectrum of the SM fermions. And, these same operators also play a
role in deriving a correct upper bound on the total proton
lifetime. Clearly, there is a prospect to use this connection to
establish an upper bound on scale of the ultraviolet completion of
our theory. We will present our findings on this issue in near
future~\cite{DFR}.

\item It is often argued that if $\Sigma_{24}$ breaks $SU(5)$ down
to the SM and $\Psi_D$ gets electroweak VEV $v$ then $\Sigma_3$
must get an induced VEV at the tree-level~\cite{Buras}. We find
this to be model dependent statement. Namely, if only $SU(5)$
invariance is imposed then one can always arrange that $\Sigma_3$
does not get a VEV at the tree-level by fine-tuning.

\end{itemize}
\section{Summary}

We have presented phenomenological aspects of the simplest
realistic $SU(5)$ model, where the Higgs sector is composed of the
${\bf 5}$, ${\bf 15}$ and ${\bf 24}$ dimensional representations.
The scenario accommodates realistic fermion masses and gauge
coupling unification. Moreover, it predicts experimentally
accessible proton decay. More specifically, there exists an upper
bound on the total proton decay lifetime $\tau_p \leq \ 1.4 \times
10^{36}$\,years. And, the bound correlates with the leptoquark
mass. Since the next generation of proton decay experiments is
expected to improve current bounds by a few orders of magnitude,
our simplest non-supersymmetric $SU(5)$ model will be certainly
tested or ruled out. The proposed scenario, if further extended
with singlet fermions and bosons, could accommodate leptogenesis
mechanism and be brought in agreement with other cosmological
observations.



\begin{thebibliography}{99}

\bibitem{Dorsner:2005fq}
  I.~Dorsner and P.~Fileviez~P\'erez,
  Nucl.\ Phys.\ B {\bf 723} (2005) 53.

\bibitem{Dorsner:2005ii}
  I.~Dorsner, P.~F.~Perez and R.~Gonzalez Felipe,
  arXiv:hep-ph/0512068.

\bibitem{GG}
H.~Georgi and S.~L.~Glashow, Phys.\ Rev.\ Lett.\  {\bf 32} (1974)
438.

\bibitem{TypeII}
  G.~Lazarides, Q.~Shafi and C.~Wetterich,
  Nucl.\ Phys.\ B {\bf 181} (1981) 287.
  R.~N.~Mohapatra and G.~Senjanovi\'c,
  Phys.\ Rev.\ D {\bf 23} (1981) 165.

\bibitem{Ellis:1979fg}
  J.~R.~Ellis and M.~K.~Gaillard,
  Phys.\ Lett.\ B {\bf 88} (1979) 315.

\bibitem{Georgi:1979df}
  H.~Georgi and C.~Jarlskog,
  Phys.\ Lett.\ B {\bf 86} (1979) 297.

\bibitem{Dorsner:2006dj}
  I.~Dorsner and P.~F.~Perez,
  arXiv:hep-ph/0606062.

\bibitem{Giveon}
  A.~Giveon, L.~J.~Hall and U.~Sarid,
  Phys.\ Lett.\ B {\bf 271} (1991) 138.

\bibitem{Dorsner:2004xa}
  I.~Dorsner and P.~Fileviez~P\'erez,
  Phys.\ Lett.\ B {\bf 625} (2005) 88.

\bibitem{Giveon:1991zm}
  A.~Giveon, L.~J.~Hall and U.~Sarid,
  Phys.\ Lett.\ B {\bf 271} (1991) 138.

\bibitem{Eidelman:2004wy}
  S.~Eidelman {\it et al.}  [Particle Data Group],
  Phys.\ Lett.\ B {\bf 592} (2004) 1.

\bibitem{Aoki:2004xe}
  Y.~Aoki  [RBC Collaboration],
  Nucl.\ Phys.\ Proc.\ Suppl.\  {\bf 140} (2005) 405.

\bibitem{Pati:1974yy}
  J.~C.~Pati and A.~Salam,
  Phys.\ Rev.\ D {\bf 10} (1974) 275.

\bibitem{DeRujula:1980qc}
  A.~De Rujula, H.~Georgi and S.~L.~Glashow,
  Phys.\ Rev.\ Lett.\  {\bf 45} (1980) 413.

\bibitem{Georgi:1980pw}
  H.~Georgi, S.~L.~Glashow and M.~Machacek,
  Phys.\ Rev.\ D {\bf 23} (1981) 783.

\bibitem{Barr:1981qv}
  S.~M.~Barr,
  Phys.\ Lett.\ B {\bf 112} (1982) 219.

\bibitem{Dorsner:2004jj}
  I.~Dorsner and P.~Fileviez Perez,
  Phys.\ Lett.\ B {\bf 606} (2005) 367.

\bibitem{Nakamura:2003hk}
  K.~Nakamura,
  Int.\ J.\ Mod.\ Phys.\ A {\bf 18} (2003) 4053.

\bibitem{Spergel:2006hy}
  D.~N.~Spergel {\it et al.},
  arXiv:astro-ph/0603449.

\bibitem{Fukugita:1986hr}
  M.~Fukugita and T.~Yanagida,
  Phys.\ Lett.\ B {\bf 174} (1986) 45.

\bibitem{Hambye:2005tk}
  T.~Hambye, M.~Raidal and A.~Strumia,
  arXiv:hep-ph/0510008.

\bibitem{seesaw}
  P.~Minkowski,
  Phys.\ Lett.\ B {\bf 67} (1977) 421 ;
  T. Yanagida, in {\it Proceedings of the Workshop on the Unified Theory
   and the Baryon Number in the Universe}, eds. O. Sawada et al., (KEK
   Report~79-18, Tsukuba, 1979), p.~95;
  M. Gell-Mann, P. Ramond and R. Slansky,
   in {\it Supergravity}, eds. P. van Nieuwenhuizen et al.,
   (North-Holland, 1979), p.~315;
  S.L. Glashow, in {\it Quarks and Leptons}, Carg\`ese, eds. M. L\'evy et al.,
(Plenum, 1980), p. 707;
  R.~N.~Mohapatra and G.~Senjanovi\'c,
  Phys.\ Rev.\ Lett.\  {\bf 44} (1980) 912.

\bibitem{Hambye:2003ka}
  T.~Hambye and G.~Senjanovic,
  Phys.\ Lett.\ B {\bf 582} (2004) 73
  [arXiv:hep-ph/0307237].

\bibitem{Hambye}
T.~Hambye, private communication.

\bibitem{Guth:1980zm}
  A.~H.~Guth,
  Phys.\ Rev.\ D {\bf 23} (1981) 347.

\bibitem{Linde:1981mu}
  A.~D.~Linde,
  Phys.\ Lett.\ B {\bf 108} (1982) 389.

\bibitem{Albrecht:1982wi}
  A.~Albrecht and P.~J.~Steinhardt,
  Phys.\ Rev.\ Lett.\  {\bf 48} (1982) 1220.

\bibitem{Shafi:1983bd}
  Q.~Shafi and A.~Vilenkin,
  Phys.\ Rev.\ Lett.\  {\bf 52} (1984) 691.

\bibitem{Shafi:2006cs}
  Q.~Shafi and V.~N.~Senoguz,
  Phys.\ Rev.\ D {\bf 73} (2006) 127301.

\bibitem{McDonald:1993ex}
  J.~McDonald,
  Phys.\ Rev.\ D {\bf 50} (1994) 3637.

\bibitem{DGF}
  I.~Dorsner, P.~Fileviez~P\'erez and G.~Rodrigo, in preparation.

\bibitem{Buras}
  A.~J.~Buras, J.~R.~Ellis, M.~K.~Gaillard and D.~V.~Nanopoulos,
  Nucl.\ Phys.\ B {\bf 135} (1978) 66.

\end{thebibliography}
\end{document}